\documentclass{ceurart}
\usepackage{listings}
\usepackage{tikz}
\usepackage{booktabs}
\usepackage{graphicx}
\DeclareUnicodeCharacter{2192}{$\rightarrow$}
\DeclareUnicodeCharacter{2194}{$\leftrightarrow$}
\DeclareUnicodeCharacter{2248}{$\approx$}
\DeclareUnicodeCharacter{2265}{$\geq$}
\DeclareUnicodeCharacter{2264}{$\leq$}
\DeclareUnicodeCharacter{00D7}{$\times$}
\providecommand{\tightlist}{\setlength{\itemsep}{0pt}\setlength{\parskip}{0pt}}

\let\origtextunderscore\_
\renewcommand{\_}{\origtextunderscore\allowbreak}
\makeatletter
\def\verbatim@font{\normalfont\ttfamily\small}
\makeatother
\usepackage{framed}
\definecolor{shadecolor}{gray}{0.94}
\let\origverbatimenv\verbatim
\let\origendverbatimenv\endverbatim
\renewenvironment{verbatim}
  {\snugshade\origverbatimenv}
  {\origendverbatimenv\endsnugshade}

\begin{document}

\copyrightclause{Copyright for this paper by its authors. Use permitted under Creative
  Commons License Attribution 4.0 International (CC BY 4.0).}

\conference{Semantic Shields 3: The 3rd International Workshop on Modeling for
  Cybersecurity, co-located with JOWO 2026 / FOIS 2026, September 21--22, 2026,
  Vit\'oria, Brazil.}

\title{What Must Be Done Here? Using a Bounded Application-Security Ontology at the
  Point of Change}

\author[1]{Pedro Farinha}[%
  orcid=0009-0001-0569-9020,
  email=pedro.farinha@shiftleft.pt,
]
\address[1]{Independent Researcher, Shiftleft -- Secure Software Engineering, Lda., Portugal}

\begin{abstract}
The security guidance that bears on code is real but scattered. It lives
across secure-development frameworks, verification standards, controls
catalogues, threat taxonomies, and, increasingly, normative regimes such
as the EU's NIS2 --- and most of it is written for readers other than
the engineer, or the AI assistant, at a point of change. Existing
cybersecurity ontologies largely model attacks, indicators, and risk for
detection and analysis, not the engineering-practice substance one must
act on. This paper presents a use case for an existing bounded
application-security ontology, AppSec Core, whose construction and
validation are reported separately. We show how it is \emph{used}. It
makes the engineering-actionable fraction of those authorities
\emph{addressable} as a typed \emph{working set}: the control objectives
that apply, the practices that realize them, the mechanisms that
implement them, and the evidence artifacts they expect, each traceable
to the sources that justify it. We reach that working set from two entry
points that recur in practice. \emph{Forward}, from a code change: a
change to authentication middleware scopes to its governing domain slice
and returns the obligations the change must satisfy. \emph{Inverse},
from a normative obligation: ``make this NIS2-compliant'' resolves a
generic legal duty into the concrete engineering obligations it implies
across slices, surfacing precisely the fraction of a mostly-governance
regulation that lands in code. The same working set is what a human
engineer reads and what an LLM-grounded assistant is handed as
grounding. The use turns ``read every applicable authority'' into
``satisfy this typed working set'', for a code change or a compliance
duty alike. The contribution is the demonstrated use; the artefact, its
validation, and its governance are companion work.
\end{abstract}


\maketitle

\section{Introduction}\label{introduction}

Software engineers increasingly work at points of change --- a pull
request, a new dependency, a refactor --- where a local decision has
security consequences. The principle of \emph{security by design} holds
that those consequences should be addressed as the system is built, not
retrofitted afterwards; in practice this resolves, at each point of
change, into a concrete question: \emph{what must be done here for this
change to be secure?} Answering it well requires knowing which security
controls apply to this change, how to realize them, and what evidence
shows they hold.

The guidance needed to answer exists, but it is hard to use at the point
of change. It is spread across secure-development frameworks,
verification standards, controls catalogues, maturity models, and threat
taxonomies, each with its own vocabulary and structure. An engineer
cannot read every authority for every change. A capable LLM, by
contrast, has in principle ingested these authorities, and --- well
prompted --- can recite and even apply them. But having read the corpus
is not the same as answering, for \emph{this} change, \emph{which}
obligations apply, \emph{on whose authority}, and whether the returned
set is \emph{complete}. Without a structure to resolve against, the
model offers a plausible subset, not a bounded, source-traceable one.
What is missing is therefore not more guidance, nor a model that has
read it, but a structure that makes the relevant guidance
\emph{addressable} --- and answerable in full --- from the point of
work.

Cybersecurity ontologies are a natural candidate for that structure, but
most existing ones are aimed elsewhere. Threat-intelligence ontologies
such as UCO \cite{UCO} and STUCCO \cite{STUCCO} integrate attack and
indicator schemas (CVE, CWE, CAPEC, STIX) for detection and intelligence
fusion. Foundational security ontologies such as COVER \cite{COVER} and
ROSE \cite{ROSE}, built on UFO, model security and risk concepts with
categorial depth for conceptual clarity. Both are valuable and
complementary, but neither is bounded to the security-by-design question
above: for a change at a given point in development, what must be done,
what realizes it, and what evidences it.

AppSec Core is a bounded application-security ontology built for exactly
that question. Its purpose is to extract, from heterogeneous source
documents, the substance that bears on building software --- and to make
it addressable where development happens, by a human engineer or,
increasingly, an AI coding assistant that needs grounding rather than
priors. Its design, OWL 2 DL formalization, SHACL validation, FAIR
assessment, and change-governance protocol are reported in a companion
artefact paper \cite{AppSecCore}; the design-science cycle that
populated it from thirty-one sources, and the source→entity substrate
that cycle produced, are reported in a second companion
\cite{Substrate}. We re-present none of this. This paper asks the
practical question the Semantic Shields community foregrounds:
\textbf{how is such an ontology \emph{used}?}

We answer by showing the same object --- a typed, source-traceable
\emph{working set} --- reached from two entry points that recur in
practice. \textbf{Forward}, from a code change: a change to
authentication middleware scopes to its governing slice and returns what
that change must satisfy (§3--4). \textbf{Inverse}, from a normative
obligation: a duty to comply with the EU NIS2 Directive resolves into
the engineering obligations it implies across slices (§5) --- surfacing,
from a regulation that is mostly governance text, the fraction that
actually lands in code, which would otherwise stay outside the
engineer's reach. The contribution is the demonstrated use and the
lesson it carries for cybersecurity ontologies intended for engineering
practice; it is not the artefact, the cycle, a retrieval mechanism, nor
an effectiveness study.

\section{AppSec Core in brief}\label{appsec-core-in-brief}

We recap only what the use case needs; the full account is in
\cite{AppSecCore}.

AppSec Core is a \emph{bounded} application-security ontology: it
carries the \emph{engineering-practice} substance of application
security --- what must be done to the software --- distilled from
thirty-one heterogeneous authorities. These span secure-development
frameworks, verification standards, and controls catalogues, and reach
into the engineering-relevant substance of compliance and regulatory
sources as well. Its boundary is on the \emph{kind} of substance, not
the source. From each authority it keeps the part an engineer acts on in
code; the governance substance those authorities also carry --- board
reporting, vendor-risk management, regulatory examination, workforce
training --- is routed elsewhere by design. It is that engineering
fraction, normalized and typed, that lets an obligation reach the
engineer as a concrete code-level task rather than a document to
interpret. It is partitioned into ten domain \textbf{slices} (e.g.,
\emph{Identity, Access \& Session Trust}; \emph{Input/Output
Validation}; \emph{Release Promotion}), each a coherent module with its
own per-slice contract. Within the slices, four populated entity types
carry the engineering substance, in a \emph{functional} reading:

\begin{itemize}
\tightlist
\item
  \textbf{ControlObjective} --- what must hold (a security objective for
  the slice);
\item
  \textbf{Practice} --- a human/process discipline that realizes an
  objective;
\item
  \textbf{Mechanism} --- a technical control that implements it;
\item
  \textbf{Artifact} --- the evidence the objective expects.
\end{itemize}

Three relations connect the types, and \texttt{belongsToSlice} assigns
each entity to a slice:

\begin{itemize}
\tightlist
\item
  \texttt{objective\_realized\_by\_practice} --- objective→practice
  (abbreviated \texttt{realized\_by} in the listings below);
\item
  \texttt{objective\_implemented\_by\_mechanism} --- objective→mechanism
  (\texttt{implemented\_by});
\item
  \texttt{objective\_expects\_artifact} --- objective→artifact
  (\texttt{expects}).
\end{itemize}

Each instance was normalized from independent normative sources and
carries per-identifier provenance to them; that provenance is what lets
a working set resolve to the sources that justify it.

\section{The use case: a change to authentication
middleware}\label{the-use-case-a-change-to-authentication-middleware}

Consider an engineer modifying a service's \textbf{authentication
middleware} --- the component that establishes caller identity and
session trust. The change is security-sensitive: done wrong, it can
weaken authentication strength, mishandle sessions and tokens, or break
the audit trail that detects abuse.

The engineer's question --- \emph{what must be done here?} --- maps to a
domain: identity, access, and session trust. In AppSec Core this is the
slice \textbf{ASC-02 (IAT)}. Scoping the ontology to the governing slice
yields the working set the change must satisfy.

\section{From change to a typed working
set}\label{from-change-to-a-typed-working-set}

The working set is not described; it is \textbf{queried} against the
deposited ontology, in two steps: a first query lists the governing
slice's control objectives, and a second expands an objective into its
typed obligations. Scoped to the governing slice ASC-02, the first query
returns the slice's control objectives:

\begin{verbatim}
PREFIX ac:   <.../appsec-core/v1#>
PREFIX rdfs: <http://www.w3.org/2000/01/rdf-schema#>
SELECT ?objective ?label WHERE {
  ?objective a ac:ControlObjective ;
             ac:belongsToSlice ac:SliceASC02 ;
             rdfs:label ?label .
} ORDER BY ?objective
\end{verbatim}

The seven obligations a change to identity/access/session components
must meet are returned verbatim:

\begin{verbatim}
ACO-IAT-001  Authentication Strength and Identity Assurance
ACO-IAT-002  Authorization Policy Integrity and Least Privilege
ACO-IAT-003  Access Revocation and Privilege Lifecycle Integrity
ACO-IAT-004  Session and Token Trust Boundaries
ACO-IAT-005  API Caller Trust and Service Boundary Enforcement
ACO-IAT-006  Access Abuse Detection and Auditability
ACO-IAT-007  Identity and Access Control Integrity
\end{verbatim}

Expanding an objective into a concrete task is the second query, which
follows the three relations that link the types. For \emph{Access Abuse
Detection and Auditability} (ACO-IAT-006):

\begin{verbatim}
SELECT ?relation ?target ?label WHERE {
  VALUES (?relation ?p) {
    ("realized_by"    ac:objective_realized_by_practice)
    ("implemented_by" ac:objective_implemented_by_mechanism)
    ("expects"        ac:objective_expects_artifact)
  }
  ac:ACO_IAT_006 ?p ?target .
  ?target rdfs:label ?label .
}
\end{verbatim}

it returns the practice that realizes the objective, the mechanisms that
implement it, and the evidence artifacts it expects:

\begin{verbatim}
realized_by    ACP-IAT-006  Access Abuse Monitoring and Audit Trail
implemented_by ACM-IAT-003  Periodic Review and Access Audit
implemented_by ACM-IAT-006  Structured Logging and Effective Configuration Recording
expects        ACA-IAT-003  Log Stream
expects        ACA-IAT-004  Policy Report
expects        ACA-IAT-005  Alert Policy
\end{verbatim}

The engineer's task at the change point is now explicit: ensure the
middleware emits an auditable log stream, that a review mechanism
consumes it, and that an alert policy exists --- and produce those
artifacts as evidence.

\textbf{The working set is a graph, not a checklist --- and that is
where the typing earns its keep.} The \emph{Log Stream} artifact
(ACA-IAT-003) is expected by six of the slice's seven objectives (Figure
1), so satisfying it once discharges evidence for several obligations at
once; the shared dependencies are explicit in the model, not left for
the engineer to rediscover per objective.

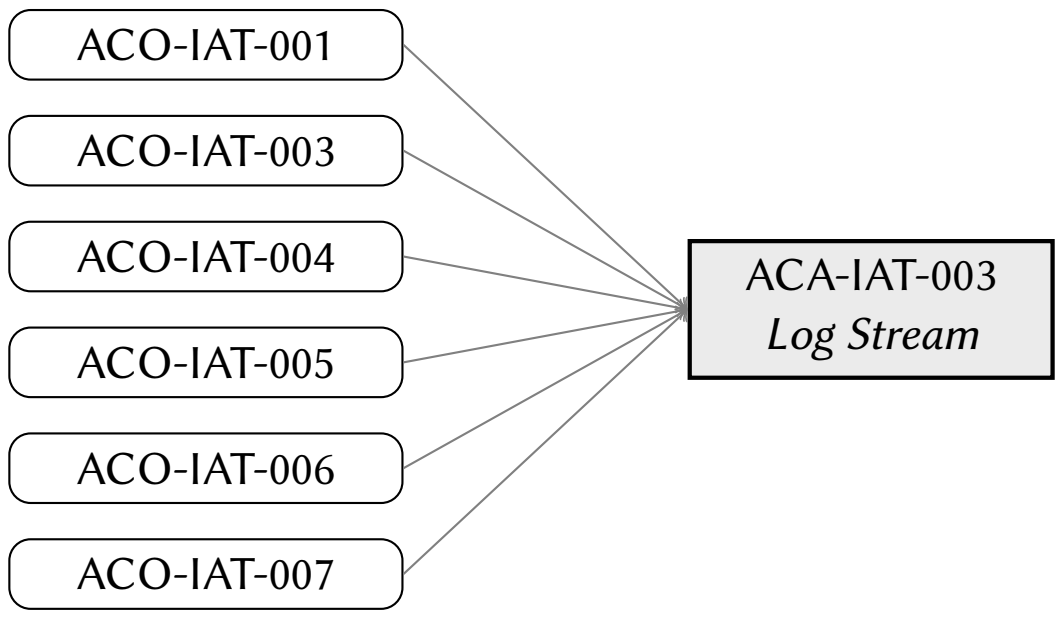
\begin{figure}[t]
\centering
\begin{tikzpicture}[font=\footnotesize,
  obj/.style={draw,rounded corners,minimum width=2.6cm,minimum height=0.46cm},
  art/.style={draw,thick,fill=black!8,align=center,minimum width=2.4cm,minimum height=0.85cm}]
  \foreach \i/\id in {1/001,2/003,3/004,4/005,5/006,6/007} {
    \node[obj] (o\i) at (0,{0.7*(3.5-\i)}) {ACO-IAT-\id};
  }
  \node[art] (ls) at (4.4,0) {ACA-IAT-003\\\emph{Log Stream}};
  \foreach \i in {1,...,6} { \draw[->,gray] (o\i.east) -- (ls.west); }
\end{tikzpicture}
\caption{The working set is a graph, not a checklist: one evidence artifact, \emph{Log Stream}, is expected by six of the IAT slice's seven control objectives. Producing it once discharges evidence for all six.}
\label{fig:graph}
\end{figure}

Each returned element carries per-identifier provenance to the deposited
source items it was distilled from, so the working set is
\emph{followable to its sources} rather than asserted --- the engineer,
or an auditor, can ask not only \emph{what} applies but \emph{on whose
authority}. Beyond that per-identifier provenance, the cycle's substrate
records which \emph{public} authorities converge on each obligation,
with a confidence band, making ``on whose authority'' inspectable
against the published standards themselves (Figure 2).

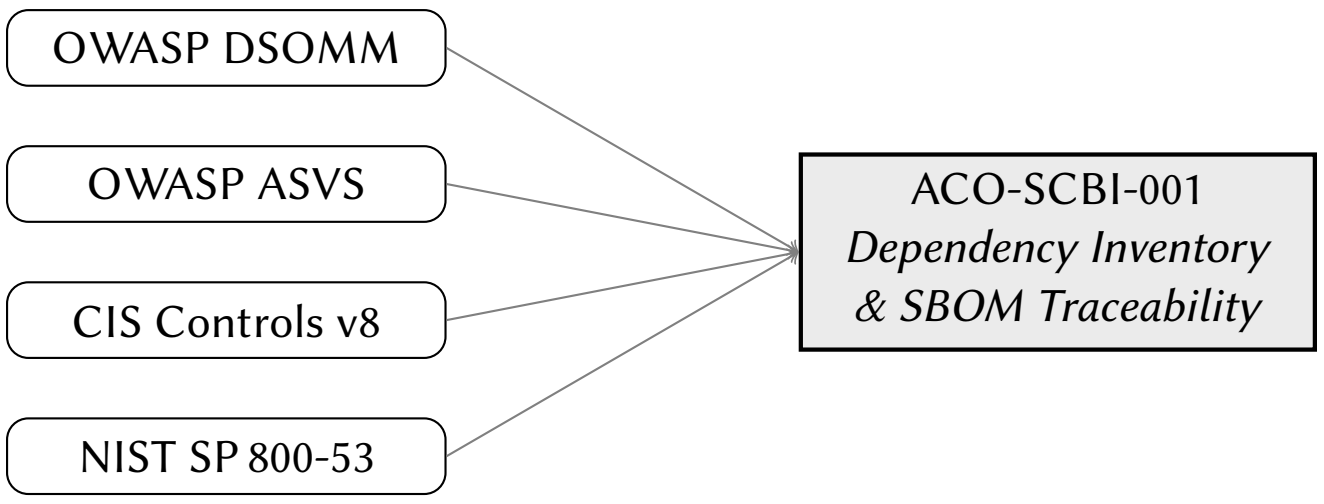
\begin{figure}[t]
\centering
\begin{tikzpicture}[font=\footnotesize,
  src/.style={draw,rounded corners,minimum width=2.9cm,minimum height=0.5cm},
  obl/.style={draw,thick,fill=black!8,align=center,minimum width=3.4cm,minimum height=1.15cm}]
  \node[obl] (o) at (5.5,0) {ACO-SCBI-001\\\emph{Dependency Inventory}\\\emph{\& SBOM Traceability}};
  \foreach \i/\name in {1/{OWASP DSOMM},2/{OWASP ASVS},3/{CIS Controls v8},4/{NIST SP\,800-53}} {
    \node[src] (s\i) at (0,{0.9*(2.5-\i)}) {\name};
    \draw[->,gray] (s\i.east) -- (o.west);
  }
\end{tikzpicture}
\caption{Ontology$\rightarrow$Sources, read from the cycle's substrate \cite{Substrate}: the
obligation ACO-SCBI-001 (\emph{Dependency Inventory and SBOM Traceability}) --- with its
practice \emph{Build-Linked SBOM Generation} and mechanism \emph{Build and Image Inventory
Generation} --- draws on several converging public authorities. This makes \emph{on whose
authority} inspectable against the published standards; the source-coverage layer is the
cycle's contribution, consumed here, not validated. Per-claim relevance scores are recorded
in the substrate.}
\label{fig:provenance}
\end{figure}

The same typed working set is what a human engineer reads and what an
LLM-grounded assistant would be handed as grounding context; such
delivery, through a Model Context Protocol (MCP) server over an authored
corpus, is companion work \cite{Retrieval} and is not part of this use
case.

The same forward query, rescoped to another slice, behaves identically:
for a change to an input-handling path it returns slice ASC-07's eight
objectives and their typed expansions, with \emph{Validation Evidence}
(ACA-IVF-003) standing in for \emph{Log Stream} as the artifact shared
across most of them. Only the governing slice differs; the workflow does
not.

\section{A second entry point: from a regulation to a working
set}\label{a-second-entry-point-from-a-regulation-to-a-working-set}

The forward use starts from code. The same working set is reachable from
the other side --- from an obligation. Suppose the driver is not a diff
but a mandate: \emph{the service must comply with the EU NIS2 Directive}
\cite{NIS2}. NIS2 Article 21 requires ``appropriate and proportionate
technical, operational and organisational measures to manage
cybersecurity risks'' --- a generic legal duty, not a code instruction.
The engineering question hidden inside it is the one from §1: \emph{what
must be done, in code, to discharge this?}

Here the query runs in the inverse direction. Rather than scoping to a
slice, it starts from the source and follows the cycle's source→entity
substrate \cite{Substrate} to the typed entities the regulation's claims
land on:

\begin{verbatim}
PREFIX ac: <.../appsec-core/v1#>
SELECT ?slice ?entity WHERE {        # over the cycle substrate
  ?item  ac:source "eu_nis2" .
  ?claim ac:item_ref ?item ;
         ac:target_core_entity ?entity ;
         ac:slice ?slice .
}
\end{verbatim}

NIS2's cyber-risk-management articles (Articles 20--23, with Article 21
chief among them) decompose, in the substrate, into 70 claims, which
land across the ontology's engineering surface as follows:

\begin{verbatim}
Threat Modeling & Risk       13     Integration Trust          6
Identity, Access & Session   10     Secret Handling & Config    5
Release & Rollback            9     Supply-chain & Build        5
Testing & Validation          9     Input/Output Validation     3
Logging & Audit Trail         8     Architecture & Boundaries   2
\end{verbatim}

The distribution is itself informative: a risk-management directive
concentrates on \emph{Threat Modeling \& Risk} and \emph{Identity,
Access \& Session Trust} --- exactly where one would expect it to land.
The mapping is semantically coherent, not uniform noise. And each
landing point is a typed obligation the engineer already knows how to
discharge from the forward use: scoped to \emph{Identity, Access \&
Session Trust}, NIS2 reaches the same objectives, practices, and
mechanisms the authentication-middleware change did (ACO-IAT-001/002,
ACP-IAT-001/002/006, ACM-IAT-001/002/006). The generic legal duty has
become a concrete, typed, slice-organized working set --- and an
engineer, or an AI assistant asked to ``make this NIS2-compliant,''
receives the code-level residue rather than a directive to read.

Two qualifications keep this honest. First, the source→entity layer is
the cycle's \emph{coverage grounding} \cite{Substrate}, not a
per-article legal interpretation. Its scores are a
\emph{relevance-triage} signal, not a verdict. Legal prose and
engineering-practice text sit in different registers, so even a
genuinely relevant obligation aligns at modest similarity; the
substrate's role is to \emph{surface} candidate obligations for a human,
or an auditor, to confirm, not to certify them. Judging which candidates
are relevant, and the systematic decomposition of each regime, is
companion work; this use consumes the existing substrate rather than
validating it.

Second, this is exactly why the count differs from the working set's
size: the substrate records the external \emph{claims} mapped into a
slice --- a \emph{claim} being one decomposed normative statement, not
the source item itself --- whereas the working set is the slice's
\emph{typed instances} (for IAT, the seven objectives with their six
practices, six mechanisms, and six artifacts). NIS2 is overwhelmingly
governance text; only its engineering-actionable fraction reaches the
ontology at all --- which is the point. The ontology extracts from a
normative document precisely the part that lands in code and leaves the
rest --- board reporting, supervision, vendor governance --- where it
belongs.

\section{What the use demonstrates}\label{what-the-use-demonstrates}

The use turns an open-ended reading task --- ``consult the relevant
chapters of every applicable authority'' --- into a bounded, checkable
one: ``satisfy this typed working set against the diff.'' What we
observed in running it, slice by slice, is what makes it work in
practice:

\begin{itemize}
\tightlist
\item
  The \emph{bounded scope} meant the query returned the governing
  slice's objectives, not an open search across all of security guidance
  --- the answer was closed and finite, and stopped at the slice that
  governs the change.
\item
  The \emph{functional typing} meant each returned item carried its kind
  of action --- a practice to adopt, a mechanism to implement, an
  artifact to produce --- so the working set read as tasks to discharge,
  not topics to read.
\item
  The \emph{provenance} meant each obligation was followable to the
  deposited source items behind it, so the engineer could see \emph{why}
  an item was required, not only \emph{that} it was.
\end{itemize}

What distinguishes the use is less the query than the \emph{shape of its
answer}. Retrieval and knowledge-graph methods typically answer by
returning \emph{more} --- ranked passages, expanded neighbourhoods,
related concepts --- and leave the engineer to judge where to stop. Here
the query \emph{stops}: scoped to the governing slice, it returns a
closed, finite working set --- every objective the ontology files under
that slice, each expanded into its typed, source-backed obligations, and
nothing outside it. That closure is structural, not incidental. Across
all ten slices every objective resolves to a practice and a mechanism
within its own slice, and no relation crosses a slice boundary; a
slice-scoped query is therefore bounded by the partition itself, not by
where the engineer chooses to stop. The three properties above ---
\emph{bounded scope}, \emph{functional typing}, \emph{per-instance
provenance} --- are together a lightweight design stance an
engineering-facing security ontology can adopt, and it is that stance,
not the query language, that turns the ontology from something to
\emph{consult} into something to \emph{satisfy}.

Shown from both entry points --- forward from a code change, inverse
from a regulation --- the result is the same kind of object: a closed,
typed, source-backed working set. That a local engineering question and
a top-down compliance duty resolve to the \emph{same} structure is the
use's most general lesson: the bounded ontology is the meeting point
where ``what must I do here?'' and ``what must I do to comply?'' become
one answer. Whether this carries to other heterogeneously regulated
domains is a question this evidence can raise but not settle.

\section{Related cybersecurity
ontologies}\label{related-cybersecurity-ontologies}

The use case sits alongside, not against, established cybersecurity
modeling. Threat-intelligence ontologies (UCO \cite{UCO}, STUCCO
\cite{STUCCO}) and cybersecurity knowledge graphs model attacks,
indicators, and countermeasures for detection and analysis; the analysis
of D3FEND in \cite{BoostingD3FEND} shows the modeling care that domain
demands. Foundational security ontologies (COVER \cite{COVER}, ROSE
\cite{ROSE}) provide categorial accounts of value, risk, and security
from which conceptual clarity follows. AppSec Core is differently aimed:
it normalizes engineering-practice substance for use at the point of
change, and combines with these families without semantic conflict --- a
threat surfaced by a threat-intelligence model, or a concept clarified
foundationally, can motivate or refine what an objective requires. The
systematic mapping in \cite{FAIRsurvey} further frames why a published,
reusable, provenance-bearing artefact matters for such use.

\section{Limitations and relation to companion
work}\label{limitations-and-relation-to-companion-work}

These are worked use cases, illustrative rather than an empirical study
of effectiveness; no comparative baseline or outcome metrics are
reported here, and whether the working set measurably improves
engineering or agent outcomes is a separate, pre-registered companion
question \cite{Eval}. For space, the full expansion into practice,
mechanisms, and artifacts is shown for one representative objective
(ACO-IAT-006); the pattern is identical for the others in the slice.

A trade-off follows from the boundedness itself. Because every relation
stays within its slice, a change confined to one domain yields a single
closed working set --- but a change that spans domains (authentication
logic that also alters the audit trail, say) must query each governing
slice in turn, and the ontology models no interaction between them. The
slice partition that makes each answer closed is also what leaves
cross-domain composition to the engineer; surfacing inter-slice
dependencies is out of scope here. The same burden appears on the
normative side: an obligation that emerges only from the joint reading
of several provisions --- common in NIS2 --- is surfaced claim by claim,
so composing interdependent provisions into one engineering requirement
likewise remains with the reader. The forward use queries the deposited
ontology directly; the inverse use additionally consumes the
source→entity substrate, whose construction, similarity calibration, and
cross-validation are the cycle paper's contribution \cite{Substrate},
referenced and not re-derived here --- and the inverse mapping is
coverage grounding, not a validated per-article legal reading. The
ontology's formalization, FAIR assessment, and change-governance
protocol are likewise reported in the artefact paper \cite{AppSecCore}.
Both artefacts and their per-identifier provenance are deposited under
open licences (identifiers in \cite{AppSecCore}, \cite{Substrate}), so
the working sets shown here are reproducible and open to scrutiny.

\section{Conclusion}\label{conclusion}

Existing cybersecurity ontologies model the threat and the risk; an
engineer --- or an AI assistant --- at a point of work also needs to
know \emph{what must be done here}, and an organization under a new
regulation needs to know \emph{what must be done to comply}. We have
shown that a bounded, functionally typed application-security ontology
answers both by returning the same object: a typed, source-traceable
working set. Forward, a change to authentication middleware scopes to
its slice; inverse, a NIS2 compliance duty resolves into the engineering
obligations it implies across slices --- the code-level fraction of a
regulation that is otherwise governance prose. The value is in the use:
a reusable, auditable answer at the point of work, consumable by a human
engineer or a grounded agent alike. We invite the cybersecurity-ontology
community to test the working-set use against further changes,
regulations, and domains.

\section*{Declaration on Generative AI}

During the preparation of this work, the author used Anthropic Claude in
order to: draft and edit text; structure and format the manuscript
(including its LaTeX sources); and run verification queries (SPARQL/RDF)
over the deposited artefacts to confirm the figures, counts, and
identifiers reported here. All factual claims, counts, and identifiers
were checked by the author against the deposited artefacts --- the
AppSec Core ontology \cite{AppSecCore} and the cycle substrate
\cite{Substrate}. After using this tool, the author reviewed and edited
the content as needed and takes full responsibility for the
publication's content.

\bibliography{references}

\end{document}